\documentclass[aps,floats,twocolumn,epsf,prl,showpacs]{revtex4-1}
\usepackage{graphicx}
\usepackage{epstopdf}
\usepackage{hyperref}
\usepackage{color}
\usepackage{braket}
\usepackage{floatrow}
\usepackage{lipsum}
\newfloatcommand{capbtabbox}{table}[][\FBwidth]
\usepackage{booktabs}

\newcommand{\ra}[1]{\renewcommand{\arraystretch}{#1}}

\definecolor{darkred}{RGB}{165,50,50}

\AtBeginDocument{ 
	\heavyrulewidth=.08em
	\lightrulewidth=.05em
	\cmidrulewidth=.03em
	\belowrulesep=.65ex
	\belowbottomsep=0pt
	\aboverulesep=.4ex
	\abovetopsep=0pt
	\cmidrulesep=\doublerulesep
	\cmidrulekern=.5em
	\defaultaddspace=.5em
}

\begin{document}

\title{Characteristic Timescales of the Local Moment Dynamics in Hund's-metals}

\author{C. Watzenb\"ock$^a$, M. Edelmann$^b$, D. Springer$^a$, G. Sangiovanni$^b$, and A. Toschi$^a$}

\affiliation{$^a$Institute of Solid State Physics, TU Wien, 1040
Vienna, Austria}
\affiliation{$^b$  Institut f\"ur Theoretische Physik und Astrophysik and W\"urzburg-Dresden Cluster of Excellence ct.qmat, Universit\"at W\"urzburg, 97074 W\"urzburg, Germany}

\date{ \today }

\begin{abstract}
	We study the characteristic timescales of the fluctuating local moments in Hund's metal systems for different degrees of correlation. By analyzing the dynamical spin susceptibility in the real-time domain we determine the timescales controlling oscillation and damping of on-site fluctuations - a crucial factor for the detection of local moments with different experimental probes. We apply this procedure to different families of iron pnictides/chalcogenides, explaining the material trend in the discrepancies reported between experimental and theoretical estimates of their magnetic moments.
\end{abstract}

\maketitle


\noindent
{\sl Introduction.} -- Our perception of the natural world is significantly shaped by the properties of the detection process considered.
One crucial aspect is the timescale of the probing mechanism: If this is larger than the typical timescale
of the phenomenon under investigation, only averaged information will be gained.  
This general statement applies to a very broad class of detectors, ranging, e.g. from the vision process in our eyes to the case of interest  for this work: the measurement of magnetic properties in correlated materials.

Here, we focus on the detection of the local magnetic moments in correlated metallic systems. 
Their proper description is, indeed, a key to understanding many-electron systems beyond the conventional band-theory framework, being central to:
Kondo physics\cite{Hewson1993, Tomczak2019},  Mott-Hubbard\cite{Imada1998,Georges1996,Hansmann2013b} or Hund-Mott\cite{DeMedici2011, Mravlje2012, Kim2017, Isidori2019, Springer2019} metal-insulator transitions, quantum criticality of heavy fermion systems\cite{Loehneysen2007,Brando2016},  magnetic and spectroscopic properties of Ni and Fe\cite{Kuebler2000,Lichtenstein2001,Hausoel2017} and of unconventional superconductors\cite{Lee2006, Chubukov2015}.

Reflecting the high physical interest, several experimental procedures are used to detect the local magnetic moments and their manifestations\cite{Blundell2001}:
measurements of static susceptibilities 
\cite{Kuebler2000,Blundell2001}, inelastic neutron spectroscopy (INS) \cite{Fishman2018}, by integrating over the Brillouin zone(BZ) \cite{Dai2015}, x-ray absorption or emission spectroscopy (XAS or XES), etc. 

Whether it is possible to obtain an accurate description of the local moments largely depends on the relation between the intrinsic timescales of the experimental probes and those characterizing the dynamical screening mechanisms at work. 
The emerging picture is typically clear-cut if the screening processes are strongly suppressed:  In Mott or 
Hund's-Mott insulating phases coherent description of the magnetic moment properties can be easily obtained in all experimental setups.
A more complex, multifaceted situation characterizes systems where well preformed magnetic moments present a rich dynamics. 
Good examples are the strongly correlated metallic regimes adjacent to a Mott metal-insulator transition, or even better, compounds displaying a Hund's metal behavior\cite{Haule2009, DeMedici2011}, such as iron pnictides and chalcogenides\cite{Chubukov2015}. 

In this work, we illustrate how to quantitatively estimate the characteristic timescales of fluctuating moments in many-electron systems within the regime of linear response.
As a pertinent example, we apply this procedure to investigate the puzzling discrepancies between  experimental and theoretical estimates of the magnetic moment size in the different families of iron pnictides/chalcogenides, clarifying the peculiar material dependence of this long-standing issue.

{\sl An intuitive picture.} -- For a transparent interpretation of our realistic calculations, we start from some heuristic considerations on 
the dynamics of the local magnetic moment $\vec{\mu} = g \frac{\mu_B}{\hbar} \vec S$ in a correlated metal.  
The relevant information is encoded in the time dependence of its correlation function
\begin{equation}
\begin{array}{rclcl}
{\mathcal F}(t)  &\equiv& \frac{1}{2} g^2 \frac{\mu_B^2}{\hbar^2} \braket{\{\hat{S}_z(t),\hat{S}_z (0)\}},
\end{array}
\label{eqn::F}
\end{equation}
where $g\cong2$ is the Land\'e factor, $\mu_{B}$ the Bohr magneton and $\hat{S}_z = \sum_\ell \hat{s}_z^\ell$ the $z$-component of the total spin moment hosted by the correlated atom (e.g., a transition metal element), built up by the unpaired electronic spins $s_z$ of its partially filled $d$ or $f$ shells \cite{Blundell2001}. We stress that Eq.~(\ref{eqn::F}) describes {\sl both} the static (thermal) and dynamic (Kubo) part of the response \cite{Wilcox1968}, which is needed for our study.
In general, one expects the maximum values of ${\mathcal F}(t)$ at $t=0$: This describes the instantaneous spin configuration of the system, often quite large in a multiorbital open shell due to the Hund's rule. Because of electronic fluctuations, the probability of finding a magnetic moment of the same size and the same orientation will be decreasing with time. At a first approximation, one can identify two distinct patterns for this process: (i) a gradual rotation (with constant amplitude) and (ii) a progressive reduction of the size of the local moment. Within this simple picture, two characteristic time (and energy) scales for the local moment dynamics are naturally defined:  (i)  the period of the rotation ($t_{\bar\omega} \! \propto \! \frac{1}{\bar\omega} $) and (ii) the characteristic time ($t_\gamma \! \propto \! \frac{\hbar}{\gamma}$) for the amplitude damping.  

The values of the characteristic timescales may vary considerably from one material to another, with overall larger values associated to a suppressed electronic mobility. 
In the extreme case of a Mott insulator, one expects to observe long-living magnetic moments, consistent with the analytic divergence of the timescales found in the fully localized (atomic) limit ($t_{\bar\omega}$, $t_\gamma \rightarrow \infty$).
On the opposite side, in a conventional (weakly correlated) metal both scales will be extremely short, roughly of the order of the inverse of the bandwidth $W$ of the conducting electrons ($t_{\bar\omega} \! \sim \! t_\gamma \! \propto  \! \frac{\hbar}{W}$).
The most interesting situation is realized in a correlated metallic context. Here, the slowing down of the electronic motion, induced by the electronic scattering, increases the values of both timescales that remain finite, nonetheless. The enhancement will depend on specific aspects of the many-electron problem considered, possibly affecting the two timescales in a different fashion: This leads to the distinct regimes of {\sl underdamped} ($ t_\gamma \gg t_{\bar\omega}$)  and  {\sl overdamped}  ($t_\gamma \ll  t_{\bar\omega}$) local moment fluctuations, schematically depicted in Fig.~1. 
The actual hierarchy of the timescales will strongly impact the outcome of spectroscopic experiments.  Further, quantitative information about the dynamics of the magnetic fluctuations at {\sl equilibrium} may also provide important information for the applicability of the adiabatic spin dynamics\cite{Sayad2015,Sayad2016, Stahl2017} and, on a broader perspective, crucial insights for the  highly nontrivial interpretation of the {\sl out-of-equilibrium} spectroscopies.


\begin{figure}[t!] 
        \centering
                \includegraphics[width=0.8\textwidth]{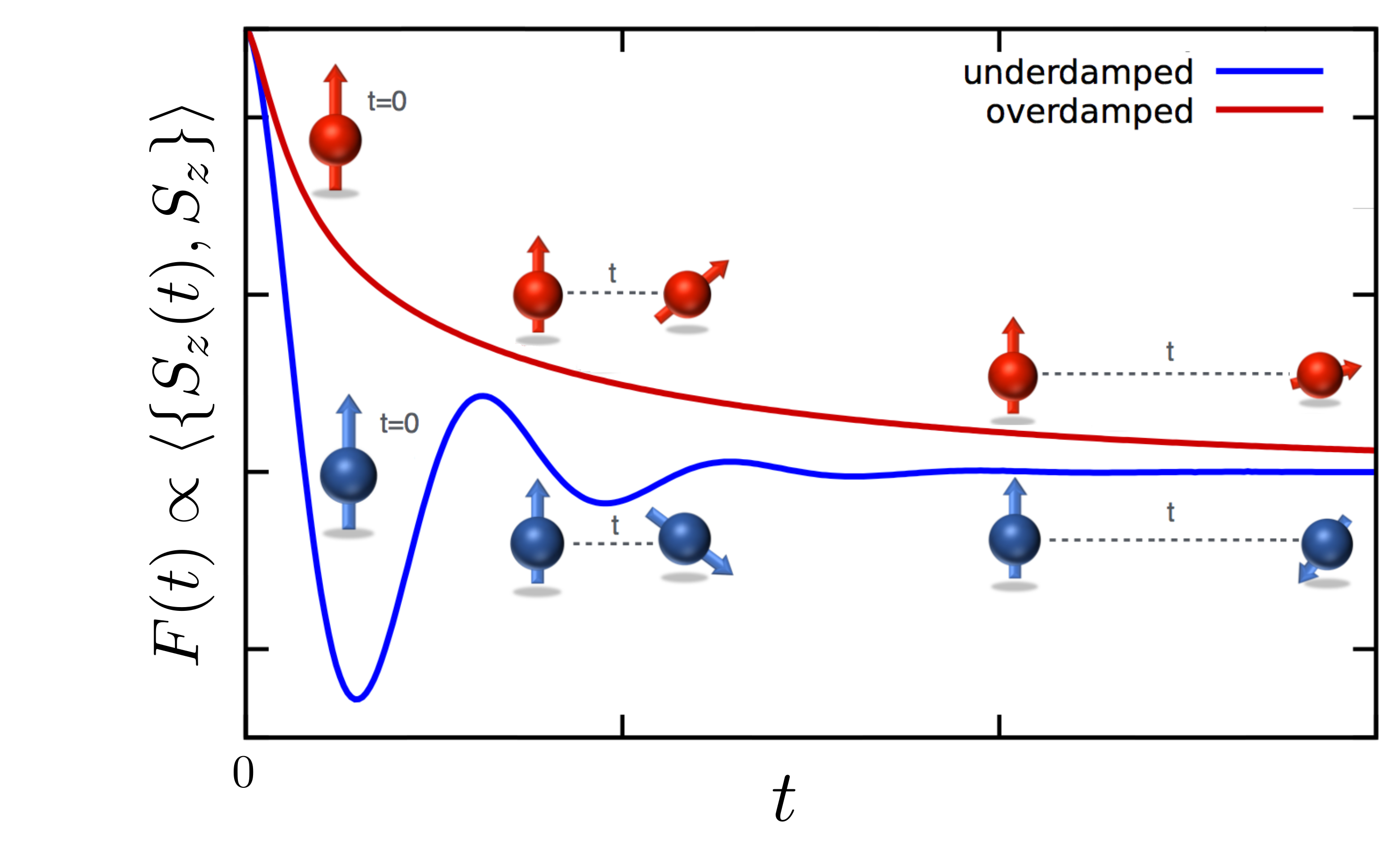}
                \vspace{-00mm}
        \caption{Schematic representation of the time decay of local spin correlations in the 
        underdamped/overdamped regimes.  \label{Fig1} }
\end{figure}
 
{\sl Quantification of timescales.} --  The procedure to quantitatively estimate the characteristic timescales from many-electron calculations and/or experimental measurements relies on the Kubo-Nakano formalism for linear response. Here, we recall that the dynamical susceptibility is defined as
\begin{eqnarray}
\label{eqn::chitau}
\chi(\tau) &\equiv& \braket{T_\tau \hat{S}_z (\tau)\hat{S}_z(0)}
\end{eqnarray}
in imaginary time ($T_\tau$ is the imaginary time-ordering operator).
The corresponding (retarded) spectral functions  $\chi^R(\omega)$ are obtained via analytic continuation of Eq.~(\ref{eqn::chitau}).
The absorption component of the spectra, Im$\chi^R(\omega)$, directly measurable (e.g.\ in INS), provides a direct route for quantifying the timescales. In particular, simple analytic expressions, directly derived for damped harmonic oscillators, can be exploited for fitting the (one or more) predominant absorption peak(s) of Im$\chi^R(\omega)$.  In the illustrative case discussed above, one has 
\begin{equation}
\mathrm{Im} \chi^R(\omega) = A  \, \frac{2 \gamma \omega}{(\omega^2 - \omega^2_{0})^2 + 4 \omega^2 \gamma^2},
\label{eqn::imchiR}
\end{equation}
where $\gamma$ and $\omega_0$ are the scales associated to the major absorption processes active in the system under consideration (with $\hbar =1$), and the constant $A$ reflects the size of the instantaneous magnetic moment. The expression is clearly generalizable to other cases,  where more absorption peaks are visible in the spectra, as a sum of the corresponding contributions \cite{SM}.

The full time-dependence of the fluctuating local moment, which will reflect the interplay of the timescales defined above, is eventually obtained via the fluctuation-dissipation theorem
\begin{equation}
\begin{array}{rcrl}
{\mathcal F}(t) &=& \frac{1}{\pi}& \int_{0}^{\infty} \mathrm{d}\omega \, \cos(\omega t) \coth(\beta/2 \omega) \,\mathrm{Im}{\chi^{R}}(\omega),
\label{eqn::fluctdiss}
\end{array}
\end{equation}
where $\beta=(k_B T)^{-1}$ is the inverse temperature.

\begin{figure*}[t!] 
\centering
\includegraphics[width=1.00\textwidth,angle=0]{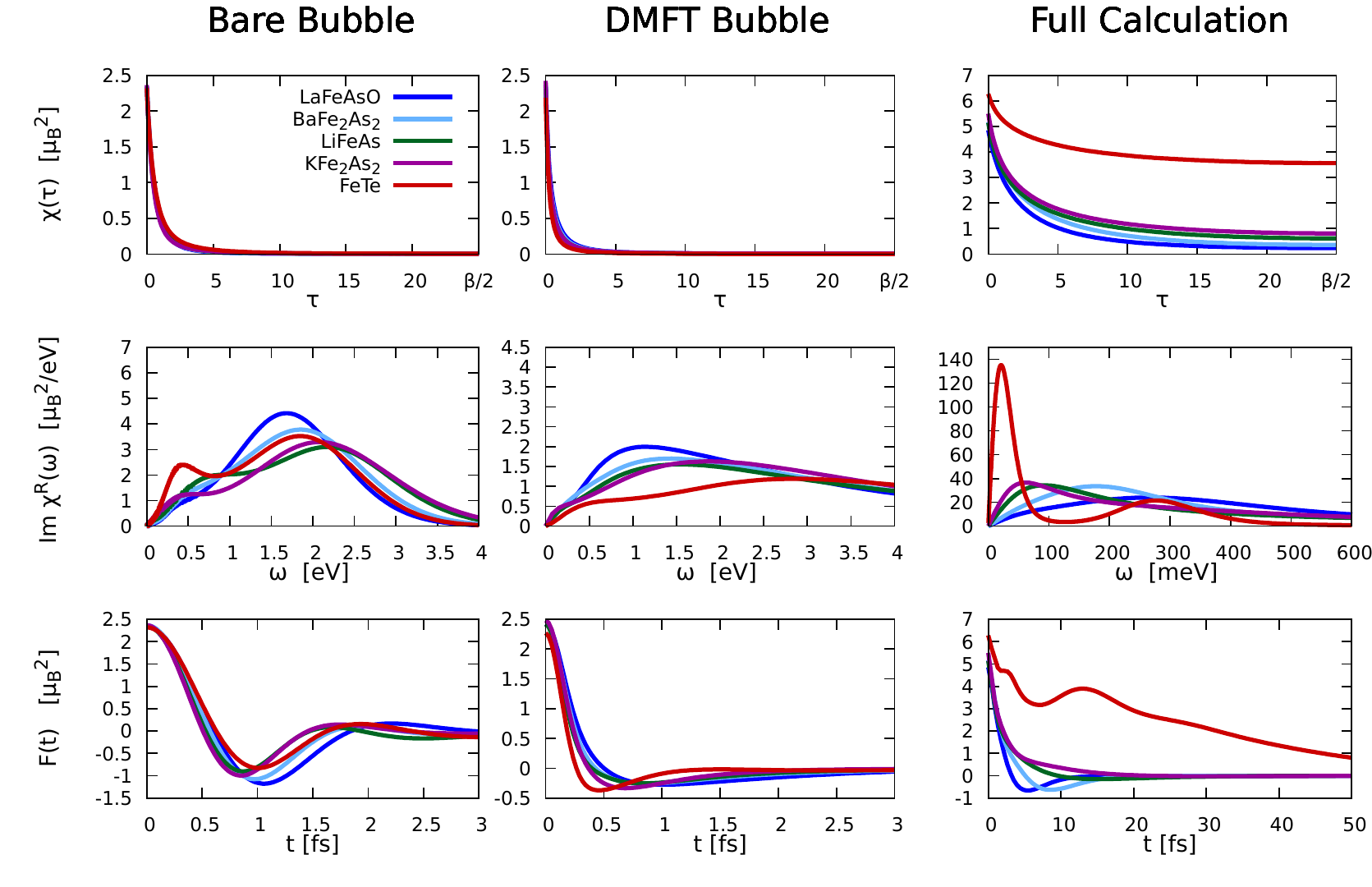}  
\caption{Spin susceptibility of the $3d$-Fe atoms as a function of imaginary time (first row), corresponding absorption spectra in real frequency (second row) and correlation function in real time (third row), computed for different families of iron pnictides or chalcogenides at $\beta=50$ eV$^{-1}$ ($T\approx232$K) in the DFT+DMFT (third column), compared with the corresponding results of the bare (first column) and the DMFT (second column) bubble calculations.}
\end{figure*}

{\sl The case of the Hund's-metals.} --  While the procedure illustrated above is applicable to all spectroscopic experiments of condensed matter systems, we will demonstrate its advantages for studying Hund's metals\cite{Haule2009,DeMedici2011}, where the dynamics of fluctuating moments is of particular interest \cite{Werner2008}.
These systems can be viewed as a new ``crossover"-state of matter, triggered by sizable values of the local Hubbard repulsion ($U$) {\sl and} Hund's rule coupling ($J$), when the corresponding atomic shell is (about) one electron away from a half-filled multiorbital configuration.  At strong coupling, the interplay between $U$ and $J$ can induce either a Mott or a charge-disproportionate Hund's insulator \cite{Fanfarillo2015,Isidori2019}. Out of half-filling, the competition between these two tendencies can also stabilize a metallic ground state in the presence of high values of the electronic interaction\cite{DeMedici2011,Fanfarillo2015,Isidori2019,Backes2015}.
The emerging physics of a {\sl large} local magnetic moment fluctuating in a strongly correlated {\sl metallic} surrounding evidently represents one of the best playgrounds to applying our time-resolved procedure. 
 
The prototypical class of materials displaying Hund's metal physics is represented by the iron pnictides or chalcogenides. 
These compounds, which often display unconventional superconducting phases upon doping, are also characterized by interesting magnetic properties \cite{Chubukov2015,Dai2015,Boeri2018}. 
Both the ordered magnetic moments (measured by neutron diffraction in the magnetically ordered phase) {\sl and} the fluctuating moments (measured by INS in the paramagnetic high-$T$ phase)  are reported to be systematically lower\cite{Tam2015} in experiment than in (static) local spin density approximation (LSDA) calculations (predicting a large ordered moment of about 2$\mu_B$ for almost all compounds of this class). It was also noted that, surprisingly, the larger discrepancies are found for the ``less correlated" families 1111 (e.g. LaFeAsO) and 122 (e.g.~BaAs$_2$O$_2$), which display milder quasiparticle renormalization effects and are characterized by lower values of the screened Coulomb interaction estimated in constrained random phase approximation (cRPA)\cite{Miyake2010}. Significantly smaller (or almost no) deviations are reported, instead, for the most correlated families such as the 11 subclass (e.g.~FeTe), where relatively large local moments are found both in neutron experiments and theory. 
Previous dynamical mean-field theory (DMFT) studies of the INS results suggested\cite{Hansmann2010, Liu2012, Toschi2012, Wang2013} that the local spin fluctuations on the Fe atom--whose time-resolved description is the  central topic here--may be responsible for the observed discrepancies. These works were restricted to one compound or (at most) one family only, and did not analyze the real-time domain. Hence, no definitive conclusion could be drawn about this issue, motivating the present computational material study.

{\sl Ab-initio + DMFT calculations.} -- We report here on our density functional theory (DFT) + DMFT calculations \cite{Kotliar2006,Held2007} of the local spin susceptibilities in the iron pnictides/chalcogenides. 
Different from preceding works, we computed the spin-spin response functions on equal footing for several different compounds, chosen as representative of the most relevant families (1111, 122, 111, 11).  As a step forward in the theoretical description, we put emphasis on a quantitative time-resolved analysis of the results, eventually allowing for a precise interpretation of the physics at play and of the spectroscopic results. 

For our DMFT calculations\cite{SM, w2dynamics}, we considered a projection on the Fe-$3d$ (maximally localized) Wannier-orbital manifold.  We assume an on-site electrostatic interaction with a generalized (orbital-dependent) Kanamori form. The corresponding Hamiltonian reads:
\begin{eqnarray}
H &=& \sum_{{\bf k} \sigma lm} \,  H_{lm}^{\phantom \dagger}({\bf k}) \; c^{\dagger}_{{\bf k} l \sigma}\, c^{\phantom \dagger}_{{\bf k} m \sigma} 
+ H_{\rm int}
\label{eqn::ham}
\end{eqnarray}
where $l,m$ are orbital indices, {\bf k} denotes the fermionic momentum, and $\sigma,\sigma'$ the spin, and
\begin{equation}
\begin{array}{rcl}
H_{\rm{int}} &=& \sum_{{\bf r} l}   U^{\phantom \dagger}_{ll} \; n_{{\bf r} l \uparrow} \, n_{{\bf r} l \downarrow} \\ &  +  & 
                       \sum_{{\bf r} \sigma \sigma', l<m} \, \left( U^{\phantom \dagger}_{lm}  \! - \! J^{\phantom \dagger}_{lm}\delta_{\sigma \sigma'} \right) \,  n_{{\bf r} l \sigma} \; n_{{\bf r} m \sigma'} \\
&- &  \sum_{{\bf r}, l \neq m} \! J^{\phantom \dagger}_{lm}  [c^\dagger_{{\bf r} l \uparrow}  c^{\dagger}_{{\bf r} l \downarrow}  c^{\phantom \dagger}_{{\bf r} m \uparrow}\, c^{\phantom \dagger}_{{\bf r} m \downarrow} \! + \! c^\dagger_{{\bf r} l \uparrow}  c^{\dagger}_{{\bf r} m \downarrow}  c^{\phantom \dagger}_{{\bf r} m \uparrow}  c^{\phantom \dagger}_{{\bf r} l \downarrow}] \\
\label{Eq:Hamiltonian_int_GK}
\end{array}
\end{equation}
where {\bf r} indicates the lattice site, and the realistic values of the screened electrostatic interactions $U_{lm}$ and $J_{lm}$ for the different materials have been taken from Ref.\cite{Miyake2010},  as detailed in \cite{SM}. The orbitally averaged values of $\bar{U}$, $\bar{J}$  range from $(2.53,0.38)$eV for LaFeAsO to $(3.41,0.48)$eV for FeTe \cite{SM}.

Our DMFT results are summarized in Fig.~2, where we show the dynamical spin susceptibility on the Fe atoms of all compounds considered in its different representations:  imaginary time in the first-row panels [cf.\ Eq.~(\ref{eqn::chitau})] which is the direct output\footnote{We compute the dynamical spin-response of the auxiliary impurity model associated to the self-consistent DMFT-solution, corresponding to the momentum-averaged one in the limit of high-connectivity/dimensions, where DMFT becomes exact\cite{Georges1996,BookPavarini2014,DelRe2020}.} of the quantum Monte Carlo (QMC) solver, real-frequency in the second row [from analytic continuations], real-time in the third row [Eq.~(\ref{eqn::F}), via Eq.~(\ref{eqn::fluctdiss})]. In all cases, we performed our analysis not only for the full DMFT calculation (third column panels), which comprises --per construction-- {\sl all} purely local effects\cite{Rohringer2012,Springer2020} of the DMFT self-energy and vertex corrections, but we also evaluate, separately, the corresponding ``bubble" terms (i.e., $\chi_0 = -\beta G G$) either computed with the noninteracting Green's function ($G = G_0$, first column) or with the DMFT one ($G = G_{\rm DMFT}$, i.e.~by including the DMFT self-energy, second column). 
\begin{table}[t!]
	\centering
	\ra{1.3}
	\hspace*{-0.0cm}
	\vspace*{4.5mm}
	\begin{tabular}{@{}lllllll@{}}\toprule
		& $\omega_0 $[eV]   & $\gamma$ [eV] &  $t_\gamma$ [fs] 	& $t_{\bar{\omega}}$ [fs] && $t_{\mathrm{1P}}$ [fs]  \\
		\cmidrule{1-5} \cmidrule{7-7}
		LaFeAsO			& \textbf{0.39}	& 0.35 & 	1.9 		 	& 3.8     & &	30.80	\\
		BaFe$_2$As$_2$	& \textbf{0.28}    & \textbf{0.28} & 	2.4 			& 15.2	  &	&	19.96	\\
		LiFeAs 			& 0.30    & \textbf{0.58} & 	7.9 			& -       &	&	12.23	\\
		KFe$_2$As$_2$ 	& 0.51    & \textbf{2.08} &  10.3			& -	  	  &	&	9.08	\\
		FeTe 			& \textbf{0.029}    & 0.022 &  29.3			& 34.8	  &	&	2.14	\\
		\bottomrule
	\end{tabular}
 	\caption{Fitting \textbf{}parameters  $\omega_0$ and $\gamma$ of the absorption peak(s) computed in DMFT with Eq.~(\ref{eqn::imchiR}) (first and second column, where the largest energy scale is marked in bold);  effective lifetime $\chi(t\rightarrow \infty)\propto \mathrm{e}^{-t/t_\gamma}$ (third column); effective oscillation period $t_{\bar{\omega}} = \hbar/\sqrt{\omega_0^2 - \gamma^2}$(fourth column) and $t_{\mathrm{1P}} = \langle \hbar/ 2 Z_i \mathrm{ Im\Sigma_i(\omega \rightarrow 0)} \rangle_{\text{all orb.}}$ (fifth column) is the effective orbital averaged one-particle lifetime for the different material considered. See \cite{SM}  for further details.}
	\label{tab:fit_params}
\end{table}

A quick glance at $\chi(\tau$) already illustrates an important finding of our work:  The different band structure of the materials  as well as their self-energies does {\sl not} generate by itself any distinguishable effects in the local moment dynamics (first two columns in Fig.~2). Instead, the definite material dependence observed is almost {\sl totally} originated by vertex corrections (third column). 

One can understand the overall material trend as follows: Instantaneous ($\tau=0$) magnetic moments of similar (and large) sizes but subjected to quite different screening effects ($\tau \rightarrow \frac{\beta}{2}$).  However, only the corresponding analysis of Im$\chi^R(\omega)$ and ${\cal F}(t)$ allows to extract clear-cut physical information. By looking at the data for $\mathcal{F}(t)$, we easily note that the moment dynamics described by the ``bubble terms" (with or without $\Sigma_{\rm DMFT}$) is controlled by very short timescales for oscillation and damping ($ \sim  0.5$ fs), roughly corresponding to $\sim \hbar/ W$.
The inclusion of vertex corrections causes, instead, a significant {\sl and} strongly material-dependent slowing down of the dynamics: In the ``least-correlated"  LaFeAsO, we  already observe oscillation and damping over {\sl few} fs (one order of magnitude {\sl larger} than in the noninteracting case). These timescales visibly increase considering more correlated families, up to the extreme case of FeTe, dominated by an extremely long decay over more than $25$ fs.

The scenario emerging from the visual inspection of ${\mathcal F}(t)$ is supported, at a  quantitative level, by the fit of the main absorption peaks of Im$\chi^R(\omega)$,
see Tab. \ref{tab:fit_params} for details. The values $t_\gamma$ and $t_{\bar\omega}$ range from 3 to 30 fs, with an overall trend which trails the progressive {\it reduction} of the quasiparticle life time ($t_{1P}$) across the different families.

 \begin{figure}[t] 
	\centering
	\includegraphics[width=0.99\textwidth,angle=0]{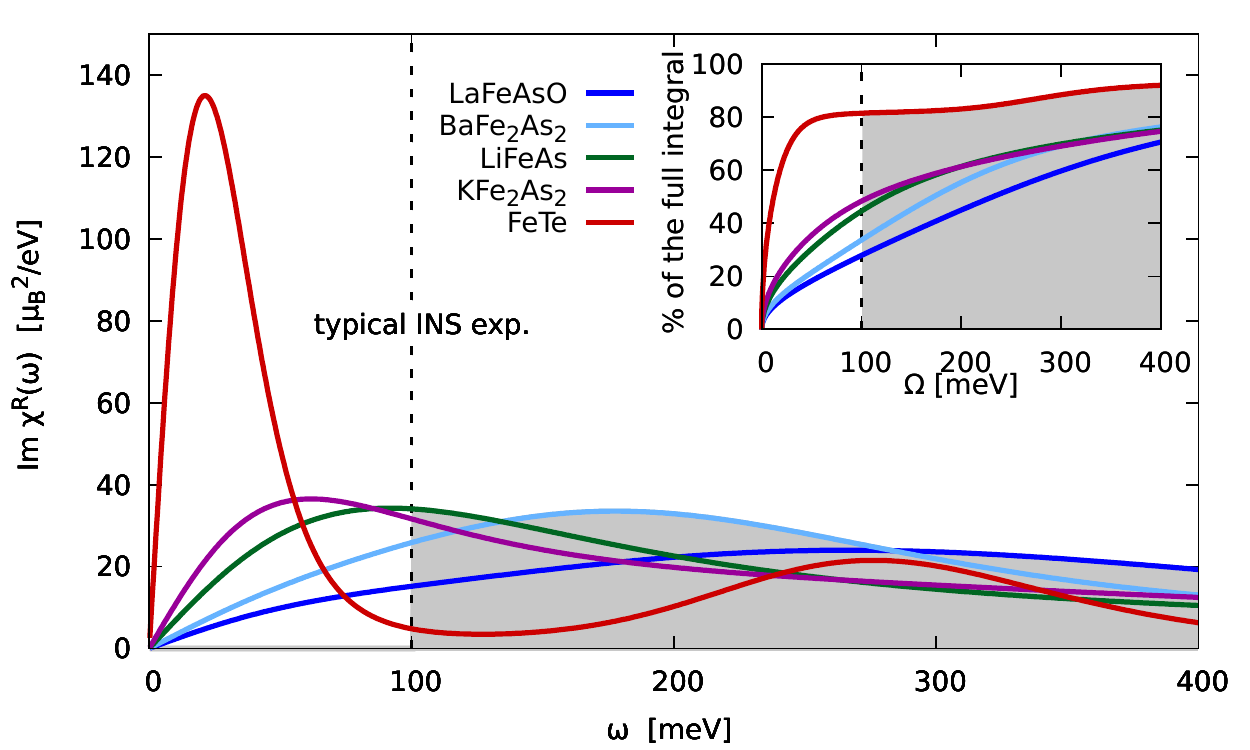}
	\caption{Material dependence of the spin-absorption spectra in the different families of the iron pnictides/chalchogenides computed in DFT+DMFT, compared with the typical  energy threshold ($\sim 100$meV) of INS experiments. Inset: Corresponding fraction of $m^2_{\rm loc}$ obtained integrating Eq.~(\ref{eqn::cutoff}) up to $\Omega$. }
\end{figure}

{\sl Spectroscopic measurements.} -- The significant spread of the estimated timescale values directly affect the detectability of the local magnetic moments ($m_{\mathrm{loc}}$) in the iron pnictides or chalcogenides. While fast probes (e.g. XAS, XES) are able to detect the high-spin instantaneous configuration of these Hund's metals, the characteristic timescale  of the INS ($t_{\rm INS} \simeq 5 \text{--} 10$ fs $\simeq \hbar / E_{\rm INS}$, with $E_{\rm  INS} = \hbar \Omega_{\rm INS} \simeq 100$meV \footnote{$t_{\rm INS}$ is defined -via the energy-time uncertainty principle- as the shortest time-interval accessible to an INS process of maximal energy $E_{\rm INS} $ of  $\mathcal{O}(100{\rm meV})$, consistent to most experiments on the materials considered. $E_{\rm INS}$ up to $300$meV was achieved in specific cases\cite{Armstrong2011,Liu2012,Wang2013}.}) are of the {\sl same order} as those in Tab. I: time-averaging effects will,  thus, lead to underestimate the local magnetic moment:
\begin{equation}
\begin{array}{rcl}
m^2_{\mathrm{loc}} &=& \frac{3}{\pi} \lim\limits_{\Omega \rightarrow \infty} \frac{ \int_{-\Omega}^{\Omega} \int_{\mathrm{BZ}} \mathrm{Im} \chi^R(\vec{q},\omega) b(\omega) \mathrm{d}\vec{q} \mathrm{d} \omega}{\int_\mathrm{BZ} \mathrm{d} \vec{q}} \\ 
&=& \frac{3}{\pi} \lim\limits_{\Omega \rightarrow \infty} \int_{-\Omega}^{\Omega} \mathrm{Im} \chi^R_{\rm loc}(\omega) \, b(\omega) \mathrm{d} \omega,
\label{eqn::cutoff}
\end{array}
\end{equation}
where $b(\omega) = 1/(\mathrm{e}^{\beta \omega} -1)$ is the Bose-Einstein distribution function (with $\hbar \! =\!1$). This is especially relevant for the ``less correlated" compounds (LaFeAsO and BaFe$_2$As$_2$), where $t_\gamma , \, t_{\bar\omega}  < t_{\rm INS}$.  In families with 
higher degrees of (e.g. for FeTe, where $t_\gamma , \, t_{\bar\omega}  > t_{\rm INS}$) the averaging effect gets ``mitigated", allowing the detection of larger magnetic moment sizes, consistent with fast probe XAS and XES experiments\cite{Kroll2008, Lafuerza2017}. 
The material dependence of local moment dynamics is directly mirrored in the progressive red shift of the first-absorption peak in Im$\chi^R(\omega)$, as shown 
in Fig.~3. Here, one can appreciate how an increasing part of the spin absorption spectra gradually enters the accessible energy window of the INS (main panel). This explains the progressively reduced discrepancies in the size of the magnetic moment (see inset) observed in the more correlated families of the iron pnictides or chalcogenides.

{\sl Conclusions.} -- We illustrated how to quantitatively investigate, on the real-time domain, the dynamics of magnetic moments in correlated systems and how to physically interpret the obtained results in terms of their characteristic timescales. 
Our procedure, exploiting the fluctuation-dissipation theorem, is then applied to clarify the results of INS experiments in several families of iron pnictides and chalcogenides. 
In particular,  the different degrees of discrepancies with respect to the standard \textit{ab initio} calculations is rigorously explained by comparing the 
 timescales of the fluctuating moments to the characteristic timescale of the INS probe. 
Remarkably, the strong differentiation among the timescales of the materials considered, crucial for a correct understanding of the underlying physics, is almost entirely due to vertex corrections. 
 
While the dynamics of the magnetic moments is particularly intriguing in the Hund's metal materials considered here, the same procedure is 
directly applicable to all many-electron systems and to fluctuations of different kinds\cite{Tomczak2019}.
A precise quantification of the characteristic timescales may provide new keys to connect the findings of equilibrium and out-of-equilibrium spectroscopies, as well as crucial information on the applicability of adiabatic spin dynamics approaches\cite{Stahl2017}. \\

\begin{acknowledgments}
\textit{Acknowledgments.} 
We thank  B.~Andersen, L.~Boeri, M.~Capone, L.~de' Medici, P.~Hansmann, K.~Held, J.~Tomczak and M.~Zingl for insightful discussions. 
We acknowledge support by the Austrian Science Fund (FWF) through Projects  SFB F41 (C.W., D.S.) and No. I 2794-N35 (A.T.) as well as by the Deutsche Forschungsgemeinschaft (DFG, German Research Foundation) through Grant No. SFB 1170, Projects No. 258499086 and No. EXC 2147, Project No. 390858490 ``ct.qmat'' (G.S.).
Calculations were performed on the Vienna Scientific Cluster (VSC).
\end{acknowledgments}

\bibliographystyle{apsrev4-1} 
\bibliography{TIMES}

\end{document}


\title{Supplemental Material for\\ ``Characteristic time-scales of the local moment dynamics in Hund's-metals''}

\author{C. Watzenb\"ock$^a$, M. Edelmann$^b$, D. Springer$^a$, G. Sangiovanni$^b$, and A. Toschi$^a$}

\affiliation{$^a$Institute of Solid State Physics, TU Wien, 1040
	Vienna, Austria}
\affiliation{$^b$  Institut f\"ur Theoretische Physik und Astrophysik, Universit\"at W\"urzburg, Am Hubland Campus S\"ud, 97074 W\"urzburg, Germany }
\maketitle

\section{Computational details}

{\sl Density Functional Theory -} For the density functional theory (DFT) calculations we employed the VASP code\cite{Kresse1993, Kresse1994}, version 5.3.3. 
As structural inputs, the experimentally found crystal structures as well as the measured lattice parameters (given in \cref{tab:structure}) have been used.\\
For all of the atoms in the given structures, we used PBE-GGA functionals.
The precise functionals used for each atom are given in \cref{tab:potentials}.
Calculations were performed on a $\Gamma$-centered MP k-mesh with $12\times12\times12$ points and $10\times10\times12$ points for I4/mmm and P4/nmm structures, respectively; The partial occupancies were calculated using the Bl\"ochl tetrahedron method.
The respective cut-off energies were, among other parameters, defined by setting the precision to HIGH, and the DOS was evaluated on 2001 points.\newline

\begin{table*}[h!]
\centering
\ra{1.3}
\begin{tabular}{@{}llllll@{}}\toprule
 Material& Crystal structure & Space group & $a$ [\AA]  & $c$ [\AA]  & z \\
\midrule
LaFeAsO				& 	ZrCuSiAs-type		  &	P4/nmm	&	4.0355\cite{possible_Lattice_Params_1111} &	8.7393\cite{possible_Lattice_Params_1111}	& 0.1418$_\text{La}$, 0.6507$_\text{As}$\cite{lattice_params_z_1111}	\\
LiFeAs 				& 	PbFCl-type tetragonal &	P4/nmm	&	3.774\cite{Lattice_Params_111}	&	6.354\cite{Lattice_Params_111}	&	0.8459$_\text{Li}$, 0.2635$_\text{As}$\cite{possible_Lattice_Params_111}	\\
BaFe$_2$As$_2$		& 	ThCr$_2$Si$_2$-type   &	I4/mmm	&	3.9625\cite{crys_struct_Ba122}	&	13.0168\cite{crys_struct_Ba122}	&0.3545$_\text{As}$\cite{crys_struct_Ba122}\\
KFe$_2$As$_2$ 		& 	ThCr$_2$Si$_2$-type   & I4/mmm	&	3.842\cite{crys_struct_K122}		&	13.861\cite{crys_struct_K122}		&	0.3525$_\text{As}$\cite{crys_struct_K122}		\\
FeTe 				&	PbO-type		      &	P4/nmm	&	3.8279\cite{lattice_parameters_ac_11}	&	6.2561\cite{lattice_parameters_ac_11}	&	0.285$_\text{Te}$\cite{lattice_parameters_z_11}\\
\bottomrule
\end{tabular}
\caption{Crystal structures for all materials under consideration. For I4/mmm materials, $c$ is given as the lattice parameter of the tetragonal cell, and $z$ in relation to this $c$.}
\label{tab:structure}
\end{table*}

\begin{table*}[h!]
\centering
\ra{1.3}
\begin{tabular}{@{}lll@{}}\toprule
 Element& Creation date & VHRFIN  \\
\midrule
La		& 	Sep 6th 2000		&	core Kr4d	\\
Fe		& 	Sep 6th 2000		&	d7s1	\\
Ba		& 	Sep 6th 2000		&	5s5p6s	\\
Te 		& 	Apr 8th 2002		&	s2p4	\\
O		& 	Apr 8th 2002		&	s2p4	\\
K 		& 	Jan 17th 2003   	&	p6s1	\\
Li 		&	Jan 17th 2003	    &	s1p0	\\
As		&	Sep 22nd 2009		&	s2p3	\\
\bottomrule
\end{tabular}
\caption{List of PAW PBE functionals used by VASP in the DFT calculations of this study. The functionals are uniquely identified by their creation date and the valence electron configuration given in the functionals by VHRFIN.}
\label{tab:potentials}
\end{table*}

{\sl Wannier Projection -} The VASP results were projected onto local orbitals via the wannier90 code\cite{Mostofi2014}.
At the time of the calculations, wannier90 integration into VAPS was only possible with wannier90 v1.2.
Specifically, all of the electronic Bloch functions in the DFT calculation were projected onto the $d$ states of Fe, no bands were marked as excluded in the wannier90.win file.
In wannier90, the k-points were identical to those from the respective VASP MP-grids.
The electronic bands of predominant Fe character are intertwined with bands of other character, such as the $p$ states of the ligands.
The degree of entanglement varied across materials, necessitating different disentanglement window parameters for wannier90, and in the case of BaFe$_2$As$_2$ and KFe$_2$As$_2$ also frozen windows.
The window positions were tweaked manually with respect to disentanglement convergence as well as agreement between original VASP bands and the bands of the Wannier Hamiltonian, the final values are given in \cref{tab:wann90_wins}.
The convergence criterion for the disentanglement as well as the wannierization was a difference in spread between successive iterations lower than $10^{-11}$.
Best results were achieved by enabling guiding centres.
The Wannier Hamiltonian served as the single-particle Hamiltonian for the Dynamical Mean Field Theory (DMFT) calculations.\newline

\begin{table*}[h!]
\centering
\ra{1.3}
\begin{tabular}{@{}lccccc@{}}\toprule
 Material& \multicolumn{2}{c}{Disentangle window [eV]} && \multicolumn{2}{c}{Frozen window [eV]} \\
 \cmidrule{2-3} \cmidrule{5-6}
  & min & max && min & max \\
\midrule
LaFeAsO				& 	-2.0473	&	2.4527	&&	&		\\
LiFeAs 				& 	-2.4441 &	2.7559	&&	&	\\
BaFe$_2$As$_2$		& 	-1.0723	&	2.7277	&&	-1.0723	&	-0.3723\\
KFe$_2$As$_2$ 		& 	-4.7175	&	3.2825	&&	-1.3175	&	3.0825		\\
FeTe 				&	-2.0831	&	2.4169	&&	&	\\
\bottomrule
\end{tabular}
\caption{Energy windows for disentangling of bands in wannier90. KFe$_2$As$_2$ and BaFe$_2$As$_2$ additionally required windows defining frozen states. Energies are given relative to the Fermi energy E$_\text{F}$=0 eV.}
\label{tab:wann90_wins}
\end{table*}

{\sl Dynamical Mean Field Theory -}  To include the effects of strong local interactions on top of the DFT, we performed DMFT simulations of an low-energy model for the entire $3d$-orbital manifold of Fe.
 The most general form of an on-site electrostatic repulsion in this manifold reads
 \begin{equation}
H_{\rm{int}} = \sum_{{\bf r} \sigma \sigma'}\sum_{lmno} \, U^{\phantom \dagger}_{lmno} \, c^\dagger_{{\bf r} l \sigma}\, 
c^\dagger_{{\bf r} m \sigma'}\,  c^{\phantom \dagger}_{{\bf r} o \sigma'} c^{\phantom \dagger}_{{\bf r} n \sigma},
\label{Eq:Hamiltonian_version2}
\end{equation}
where the full-fledged, four-indexed $U-$tensor describes the projected value of the screened Coulomb interaction on the corresponding orbital configurations.
As an {\sl ab-initio} estimate for the {\sl orbital-dependent} interaction parameters, we take the results by Miyake et al.\cite{Miyake2010}, where constrained random phase approximation (cRPA)
results for the {\sl two-orbital} interaction matrix $U_{lm}$ and $J_{lm}$ were reported for different compounds\footnote{With the only exception of KFe$_2$As$_2$, where we used the same values as for the other family-compound (BaFe$_2$As$_2$ given in \cite{Miyake2010}) as they are not available in the literature.}. Here, the $J_{lm}$ values encode the (orbital-dependent) Hund's coupling, while the $U_{ij}$  diagonal/off-diagonal matrix elements describe the inter-/intra-orbital electrostatic repulsion. 
The relation, which we exploited to extract the interaction parameters appearing in Eq.~\ref{Eq:Hamiltonian_version2},  is:
\begin{equation}
\begin{array}{rcl}
U^{\phantom \dagger}_{ijkl} = 
\begin{cases}
U^{\phantom \dagger}_{ij}, & \text{if } ijkl=ijij, \\
J^{\phantom \dagger}_{ij}, & \text{if } ijkl=iijj \text{ and }i\neq j, \\
J^{\phantom \dagger}_{ij}, & \text{if } ijkl=ijji \text{ and }i\neq j, \\
0, & \text{otherwise}.
\end{cases}
\label{Eq:Hamiltonian_index_relations}.
\end{array}
\end{equation} 
This leads to the low-energy Hamilonian used for our DMFT calculations
\begin{equation}
H = \sum_{{\bf k} \sigma lm} \,  H_{lm}^{\phantom \dagger}({\bf k}) \; c^{\dagger}_{{\bf k} l \sigma}\, c^{\phantom \dagger}_{{\bf k} m \sigma} + H_{\rm int},
\label{Eq:Hamiltonian}
\end{equation}
with 
\begin{equation}
\begin{array}{rcl}
H_{\rm{int}} &=& \sum_{{\bf r} l} \,  U^{\phantom \dagger}_{ll} \; n_{{\bf r} l \uparrow} \, n_{{\bf r} l \downarrow}  + 
                       \sum_{{\bf r} \sigma \sigma'} \sum_{l<m} \, \left( U^{\phantom \dagger}_{lm} - J^{\phantom \dagger}_{lm}\delta_{\sigma \sigma'} \right) \,  n_{{\bf r} l \sigma} \; n_{{\bf r} m \sigma'} \\
&& - \sum_{{\bf r}}\sum_{l \neq m} \, J^{\phantom \dagger}_{lm} \; c^\dagger_{{\bf r} l \uparrow}\, c^{\dagger}_{{\bf r} l \downarrow} \, c^{\phantom \dagger}_{{\bf r} m \uparrow}\, c^{\phantom \dagger}_{{\bf r} m \downarrow}
- \sum_{{\bf r}}\sum_{l \neq m} \, J^{\phantom \dagger}_{lm} \; c^\dagger_{{\bf r} l \uparrow}\, c^{\dagger}_{{\bf r} m \downarrow} \, c^{\phantom \dagger}_{{\bf r} m \uparrow}\, c^{\phantom \dagger}_{{\bf r} l \downarrow}
\label{Eq:Hamiltonian_int_GK}.
\end{array}
\end{equation} 
Physically, this corresponds to an {\sl orbital-dependent  Kanamori} interaction, where one can easily recognize  an orbital-dependent pair-hopping (the first term in the second line)  and spin-flip contribution (second term). 
 In fact, \cref{Eq:Hamiltonian_int_GK} can be regarded as an orbital-dependent generalization of the  Kanamori interaction, since for the special cases of no-orbital dependence e.g. averaged interaction parameters (where $U^{\phantom \dagger}_{ll} = U $, $U^{\phantom \dagger}_{l\neq m} = V $ and $J^{\phantom \dagger}_{l m} = J $) we recover the usual expression of the Kanamori Hamiltonian: 

\begin{equation}
\begin{array}{rcl}
H^K_{\rm{int}} &=& U^{\phantom \dagger} \; \sum_{{\bf r} l} \; n_{{\bf r} l \uparrow} \, n_{{\bf r} l \downarrow}  + 
\sum_{{\bf r} \sigma \sigma'} \, \left( V - J\, \delta_{\sigma \sigma'} \right) \sum_{l<m}  \,  n_{{\bf r} l \sigma} \; n_{{\bf r} m \sigma'} \\
&& - J \, \sum_{{\bf r}}\sum_{l \neq m} \; c^\dagger_{{\bf r} l \uparrow}\, c^{\dagger}_{{\bf r} l \downarrow} \, c^{\phantom \dagger}_{{\bf r} m \uparrow}\, c^{\phantom \dagger}_{{\bf r} m \downarrow}
- J \; \sum_{{\bf r}}\sum_{l \neq m} \, c^\dagger_{{\bf r} l \uparrow}\, c^{\dagger}_{{\bf r} m \downarrow} \, c^{\phantom \dagger}_{{\bf r} m \uparrow}\, c^{\phantom \dagger}_{{\bf r} l \downarrow}
\label{Eq:Hamiltonian_int_K}.
\end{array}
\end{equation}

To illustrate concisely the variation of the screened interaction values in the different materials, the corresponding orbitally average $U$- and $J$-values  are shown in \cref{tab:avgUJV}. 
The DMFT calculations shown in the main text were, however, performed using the orbital-resolved Hamiltonian (\ref{Eq:Hamiltonian_int_GK}).  
\begin{table*}[h!]
\centering
\ra{1.3}
\begin{tabular}{@{}lcc@{}}\toprule
			& $\overline{U}$  & $\overline{J}$ \\
\midrule 
LaFeAsO			&2.53	&0.39		\\
BaFe$_2$As$_2$	&2.81	&0.43		\\
KFe$_2$As$_2$	&2.81	&0.43		\\
LiFeAs 			&3.15	&0.43		\\
FeTe 			&3.41	&0.48		\\
\bottomrule
\end{tabular}
\caption{Average effective on-site Coulomb (U) exchange (J) interactions between two electrons on the same iron site in the $d-$ model (in eV).}
\label{tab:avgUJV}
\end{table*}

Finally, let us mention that in order to check the robustness of our conclusions, we have also performed DMFT calculations using the orbitally-averaged values for the $U$ and $J$  interaction (i.e., corresponding to a ``conventional" Kanamori interaction, not shown), finding only marginal changes to the results shown in Fig.~2 and 3 of the main text. 
Larger quantitative modifications can be found in the results of the most correlated materials, as expected, only if one neglects the spin-flip terms in Eq.~(\ref{Eq:Hamiltonian_int_K}) (e.g., when performing density-density calculations, not shown here\cite{Watzenboeck2018}). The reason is, that in this approximation one tends to overestimate the high-spin configurations in the strong-coupling regime.
  
The number of electrons in the target ($d$-) manifold was estimated directly from chemical considerations (constituent electronegativity). Throughout our calculation, we assumed that  LaFeAsO, BaFe$_2$As$_2$ LiFeAs and FeTe have a filling of $\sum_{l, \sigma} \braket{n^l_\sigma}  = 6.0$ electrons per iron atom. For KFe$_2$As$_2$ we used, instead, $\sum_{l, \sigma} \braket{n^l_\sigma}  = 5.5$ per iron atom.

To avoid double-counting of the Coulomb interaction between Fe-3d electrons already included in DFT, we used an orbital-dependent double-counting correction of the Fully-Localized-Limit (FLL) type \cite{Czyzyk1994} (adopted also for DMFT calculations of elemental Fe\cite{Hausoel2017}). The values used were determined by \cref{eqn:DCC_orbitaldependent} and are shown in \cref{tab:DCC_d}.

\begin{equation}
\begin{array}{rcl}
\mu^{FLL}_{DC}(i) &=& \mu^{FLL}_{DC}(i) + \frac{1}{4} \left(n^0 - \frac{1}{2}\right) \left( \sum_{j}(U_{ij}  - J_{ij})  \right),
\label{eqn:DCC_orbitaldependent}
\end{array}
\end{equation}

In \cref{eqn:DCC_orbitaldependent}   $ n^0 = \frac{1}{2(2l+1)} \sum_{i,\sigma} n_{i,\sigma} $ is the DFT filling and the two-indices U-matrix is related to the four-indices local (screened) Coulomb-tensor by $U_{ij}=U_{ijij}$ and $J_{ij}=U_{ijji}$ (with $i\neq j$).

\begin{table*}[h!]
\centering
\makebox[\textwidth][c]{
\begin{tabular}{@{}lccccc@{}}\toprule
            & LaFeAsO  & BaFe$_2$As$_2$ &  LiFeAs &  KFe$_2$As$_2$ & FeTe    \\
\midrule
$3z^2-r^2$	& 9.5295   &10.8510         &12.5740	& 9.8805         &13.3397  \\  
$xz$		    & 8.9790   &9.9117          &11.6900	& 9.0234         &11.9597  \\
$yz$		    & 8.9790   &9.9117          &11.6900	& 9.0234         &11.9597  \\ 
$x^2-y^2$	  & 8.2157   &9.4147          &11.5623	& 8.5699         &13.2400  \\ 
$xy$		    & 9.6732   &10.6073         &12.2523	& 9.6586         &13.5735  \\ 
\bottomrule
\end{tabular}
}
\caption{Orbital dependent double counting correction (DCC) in the fully-localized limit.}
\label{tab:DCC_d}
\end{table*}

The DMFT simulation was performed with a continuous-time quantum Monte Carlo (QMC) algorithm implemented in the code package \texttt{w2dynamics}\cite{w2dynamics}. All calculations were done at $\beta=50$[$\mathrm{eV}^{-1}$] corresponding to approximately  $232.1$K. At this temperature all the materials are experimentally found to be in the paramagnetic phase\cite{Wang2010, Li2009, Grinenko2012}. To achieve convergence in the DMFT cycle with orbital dependent Kanamori interaction for the different materials we first converged the DMFT cycle without the pair-hopping term in Eq.~(\ref{Eq:Hamiltonian_int_GK})). This was achieved by performing 30$\div$70 DMFT steps with low statistics (\texttt{Nmeas}=$10^4$, where \texttt{Nmeas} is then number of QMC measurements, see \cite{w2dynamics} for details). 
Up to 100 additional DMFT-steps were performed with higher statistics. 

For each calculations, the final convergence of the DMFT self-consistency was tested for the one-particle quantities encoded in the self-energy $\Sigma^l(\omega)$,  with respect to the previous five iterations. For the number of steps between measurements (where the minimum value gives a measure of auto-correlation time) we found a value of \texttt{Ncorr}=$1500\div2000$ to be sufficient, in line with the estimate of the average ``renewal time'' of the fermionic trace given in Ref.\onlinecite{w2dynamics}.

On top of the converged one-particle quantities we then calculated the spin-spin susceptibility in imaginary time through a single DMFT step with fixed chemical potential and one-particle properties, using \texttt{Nmeas}$>5\cdot10^4$. As a QMC-sampling algorithm we applied the recently developed States-Sampling \cite{Kowalski2019}. The result is show in the upper-right part of Figure 2 of the main text.

To obtain the DMFT-bubble susceptibility we used $\chi^{\mathrm{Bubble-DMFT}}(\tau) = \sum_l G^l_{\mathrm{loc}}(\tau) G^l_{\mathrm{loc}}(\beta-\tau)$, and for the bare-bubble we set the interaction as well as the double counting correction in the DMFT calculation to zero ($U=J=V=\mu_{DC}=0$), corresponding to $G_\mathrm{loc}=G_0$. 

\begin{figure}[h]
	\centering
	\includegraphics[width=0.75\textwidth]{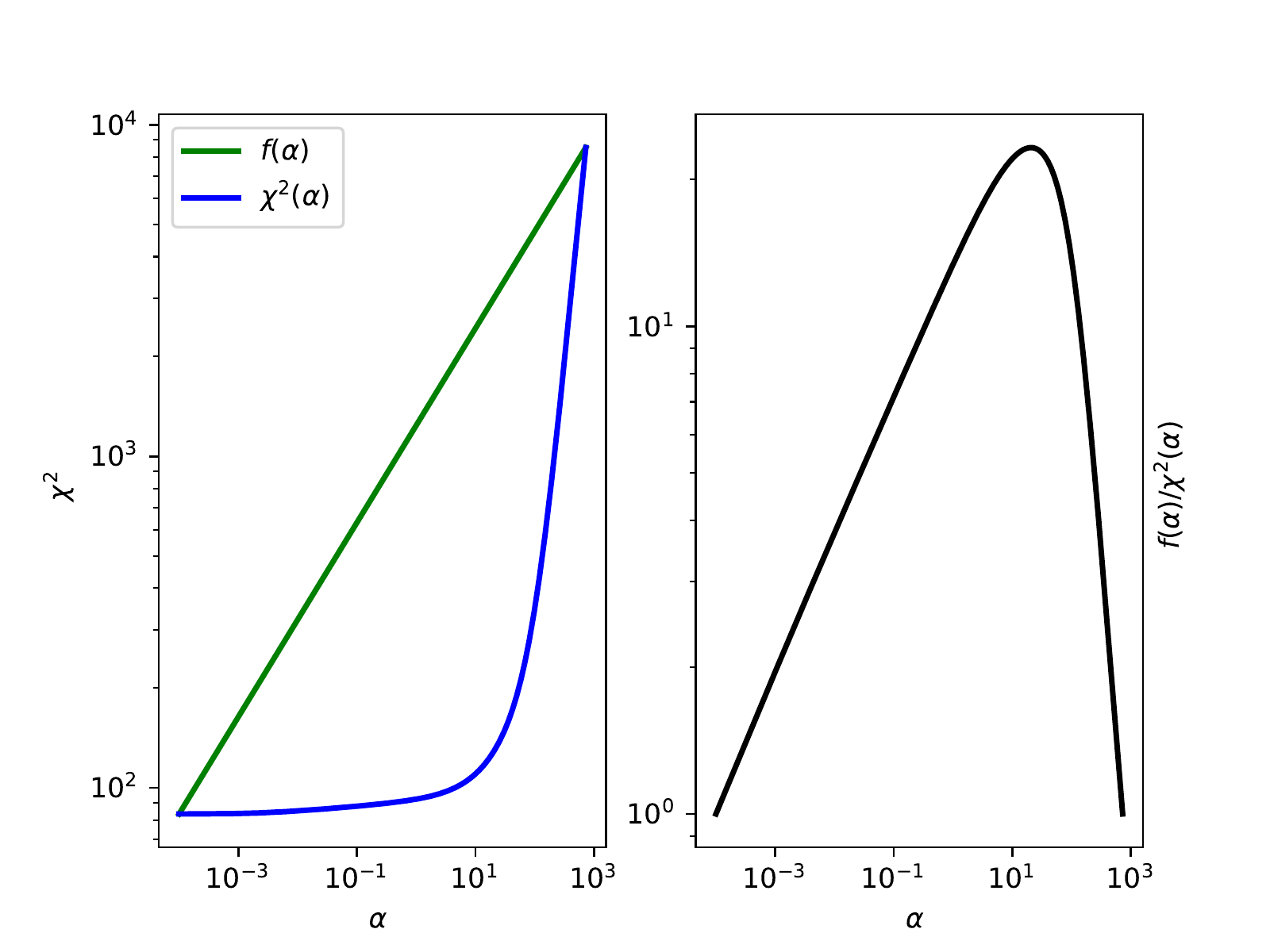}
	\caption{ Log-log-plot of the quadratic difference between the data and the fit $\chi^2$ over the entropy parameter $\alpha$ for LaFeAsO. We find overfitting (underfitting) of the data to start at $\alpha<10^0$  ($\alpha>10^2$). The spectrum corresponding to the maximum of $f(\alpha)/\chi^2(\alpha)$ (at $\alpha=2 \cdot 10^1$) could be regarded as a good analytical continuation.}  
	\label{fig:kink}
\end{figure}

{\sl Analytical continuation -} Analytical continuation from imaginary time (where the QMC data was obtained) to real frequencies was performed with the Maximum Entropy Method (MaxEnt) with the code package \texttt{Maxent} \cite{Levy2017}. This way we obtained the imaginary part of the retarded susceptibility $\chi^R(t) \equiv \frac{\mathrm{i}}{\hbar} \theta(t) \braket{[\hat{S}_z (t),\hat{S}_z(0)]}$.
The effect of different default-models (Flat, Gaussian, Lorentzian) was tested and found to be small. We chose a broad featureless Lorentzian-default model with a width of $\gamma_{\mathrm{Model}}=0.5$. Model-details are found elsewhere \cite{Levy2017}. The optimal $\alpha-$parameter (weight of the entropy term in MaxEnt) was determined by the maximum of the curvature of $\chi^2(\alpha)$. (See \cref{fig:kink}.)  This way we could reliably determine the region where neither the data was over-fitted nor the default model was take into account too strongly. One advantage of this method is invariance of the final spectrum under re-scaling the error by a global factor\cite{Watzenboeck2018}. A similar approach was already applied in \cite{Otsuki2017}. 

\newpage
{\sl One particle time scales}- For estimating one particle timescales we assumed that (in the presence of a well defined quasi-particle excitation) the one-particle Green's function $G^i(t)$ for each orbital $i$ decays in the following way: $\left| G^i(t)\right|^2  = Z_i^2 \mathrm{e}^{- \frac{t \, 2\, Z_i \, \mathrm{Im}  \Sigma_i(\omega\rightarrow 0)}{\hbar}} \propto \mathrm{e}^{- \frac{t}{t^i_{\mathrm{1P}}}}$, with $t^i_{\mathrm{1P}}\equiv \frac{\hbar}{2 Z_i \mathrm{Im}\Sigma_i(\omega \rightarrow 0)} $, where $\Sigma_i$ is the self-energy of the orbital $i$. The value of the self-energy at zero frequency as well as the orbital dependent quasi-particle mass re-normalization $Z_i=\left(1 + \mathrm d/\mathrm d \omega \mathrm{Re} \Sigma_i(\omega)\big |_{\omega\rightarrow 0}\right)^{-1}$ was extracted from the DMFT self-energy by linear interpolation of $\mathrm{Im} \Sigma(\mathrm{i}\omega_n \rightarrow 0)$ (using the Cauchy-Riemann equations for $Z_i$). The one particle time scale $t_{\mathrm{1P}}$ given in the main text was then estimated as the orbital average of $t^i_{\mathrm{1P}}$:  $t_{\mathrm{1P}} = \frac{1}{5} \sum_i t^i_{\mathrm{1P}}$. 

{\sl Spin-excitation time scales -} 
While the time scales of spin-excitations in iron-based superconductors are determined by an intricate interplay of kinetic energy (hopping) and electron-electron-interaction the main time scales can be effectively described by a much simpler model. The extraction of time scales was done by applying a uniform $\chi^2$-fit to $\mathrm{Im}\, \chi^R(\omega)$ with cutoff-values chosen for the grid such that the main-peaks structure is well within the frequency window (1eV). The cutoff excludes high-frequency data, which is usually not as well captured by MaxEnt as the low-frequency data. A variation of the cutoff by 20\% leads to a change in the time scales by less than 15\%. 

The fitting function is defined as follows: We consider the absorption spectrum of a damped harmonic oscillator, which can be obtained by the Fourier-transform of the Green's function of the differential equation $\ddot \chi(t) + 2\gamma \dot \chi(t)- \omega^2_0 \chi(t) = -\delta(t)$, i.e. $\chi(\omega) = \frac{1}{ \omega^2 - 2 \mathrm{i} \gamma \omega + \omega^2_0}$. We note that the latter has poles only on the lower half-plane, and thus it is a retarded function ($\chi(t<0)=0$). Its imaginary part (up to a proportionality-constant reflecting the material-dependent value of the unscreened local moment) defines our fitting model which, thus, reads 
\begin{equation}
 \mathrm{Im} \chi^R(\omega) = 2 \gamma \omega \frac{1}{(\omega^2 - \omega^2_0)^2 + 4 \omega^2 \gamma^2},
 \label{eqn:harm_osz_chiw}
 \end{equation}
  or correspondingly in real times

\begin{equation}	
\chi(t) = \left\lbrace
\begin{array}{lll}
\frac{\mathrm{e}^{-\gamma t}}{\sqrt{\omega_0^2 - \gamma^2  }} \sin(\sqrt{\omega_0^2 - \gamma^2}t) \theta(t)   & \phantom{aa}&\text{if } \omega_0^2 > \gamma^2\\
\frac{\mathrm{e}^{-\gamma t}}{\sqrt{\gamma^2 - \omega_0^2   }} \sinh(\sqrt{\gamma^2 - \omega_0^2 }t) \theta(t)   & \phantom{aa}&\text{if } \omega_0^2 < \gamma^2 {.}\\
\end{array}
\right.
\label{eqn:Analytical_chi_t}
\end{equation}
The asymptotic behavior, which determines the main-lifetime is given by 
\begin{equation}	
\lim\limits_{t \rightarrow \infty}\chi(t) \propto \left\lbrace
\begin{array}{llll}
\mathrm{e}^{-\gamma t} &\equiv \mathrm{e}^{-t/t^{\mathrm{under}}_\gamma}   & \phantom{aa}&\text{if } \omega_0^2 > \gamma^2\\
\mathrm{e}^{-\left(\gamma- \sqrt{\gamma^2 - \omega_0^2 }\right) t} &\equiv \mathrm{e}^{-t/t^{\mathrm{over}}_{\gamma}}  & \phantom{aa}&\text{if } \omega_0^2 < \gamma^2 .
\end{array}
\right.
\end{equation}

The corresponding parameters obtained by fitting the DMFT spectra are summarized in \cref{tab:fit_params}.

\begin{figure}[H]
	\minipage{0.32\textwidth}
	\includegraphics[width=\linewidth]{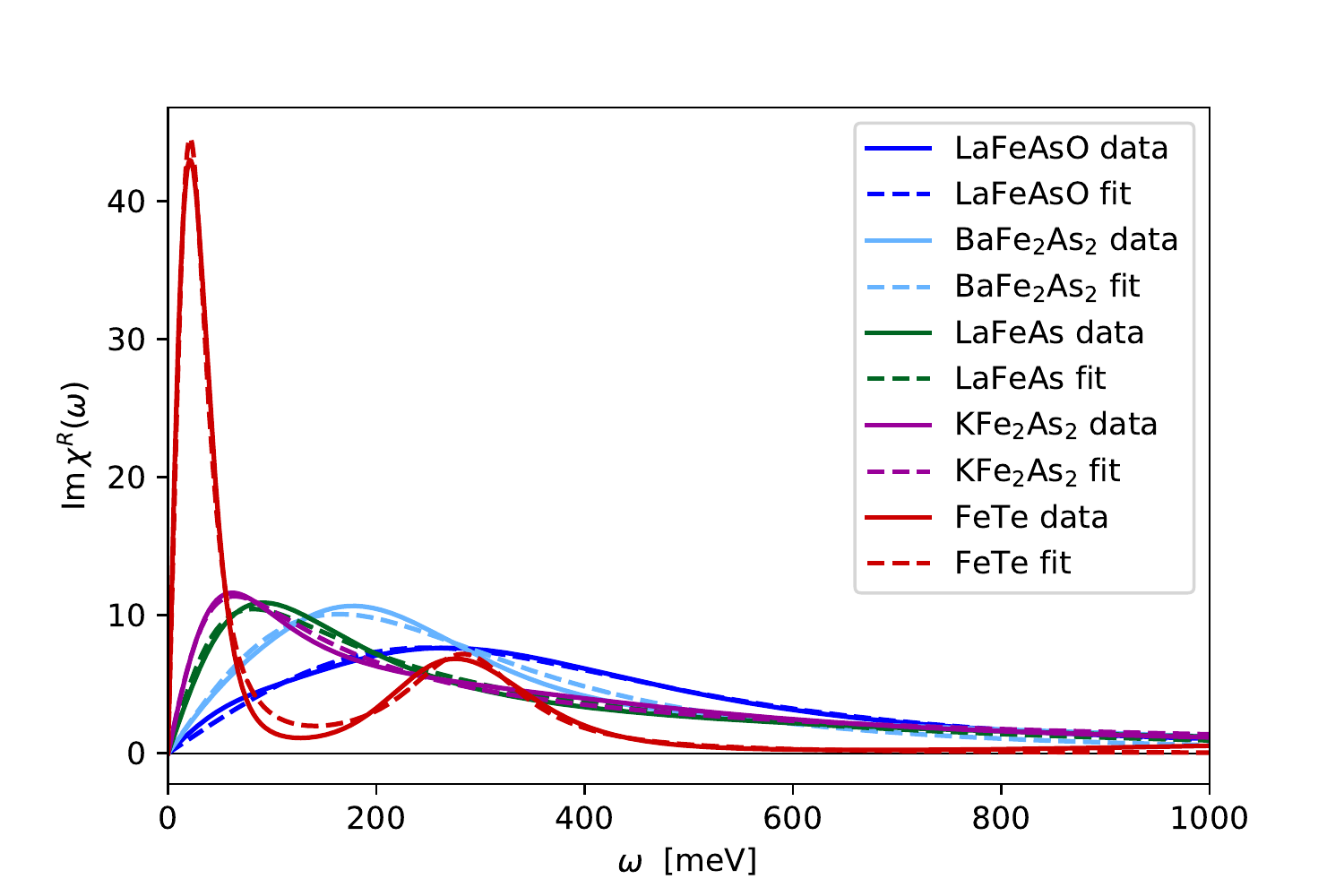}
	\endminipage\hfill
	\minipage{0.32\textwidth}
	\includegraphics[width=\linewidth]{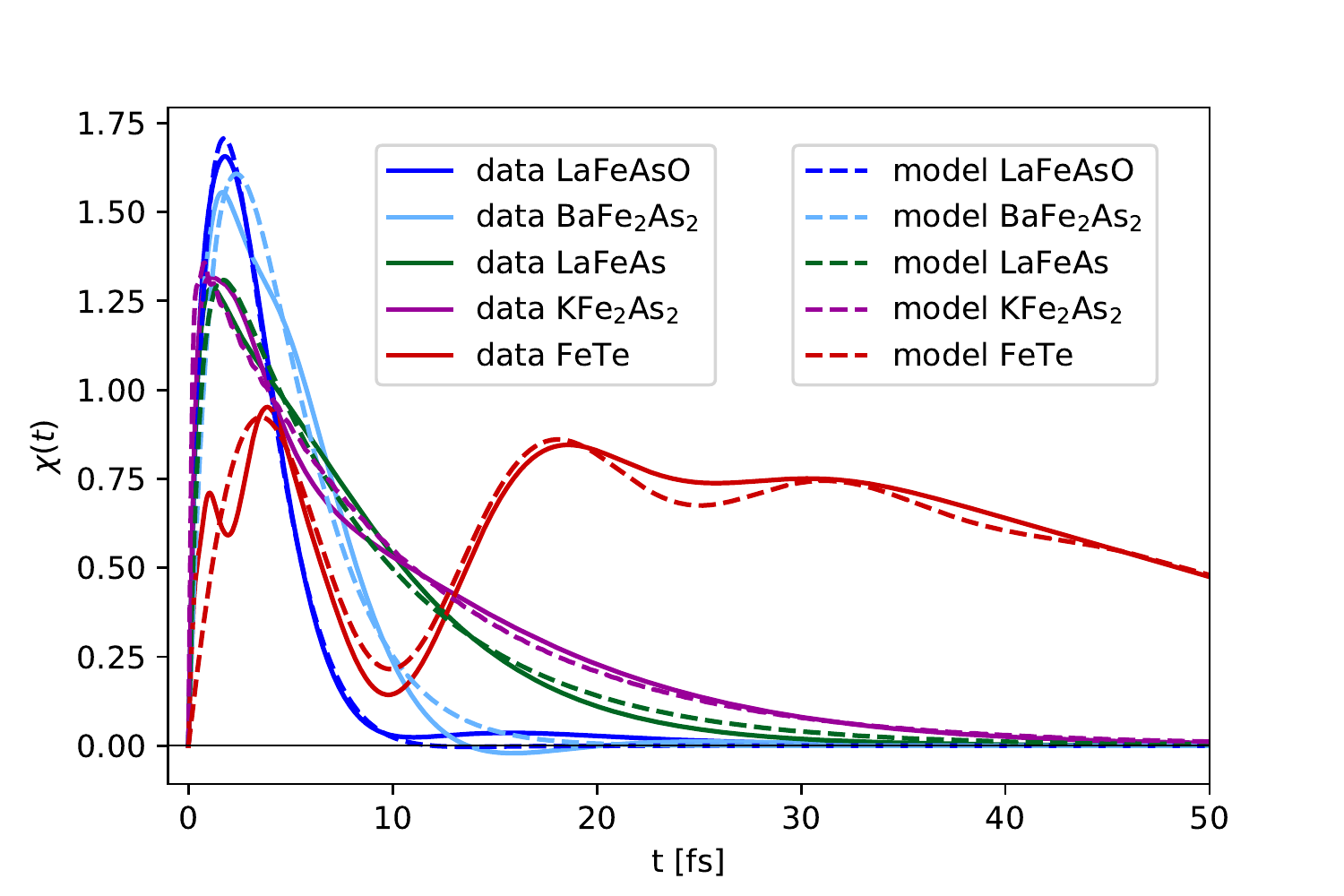}
	\endminipage\hfill
	\minipage{0.32\textwidth}%
	\includegraphics[width=\linewidth]{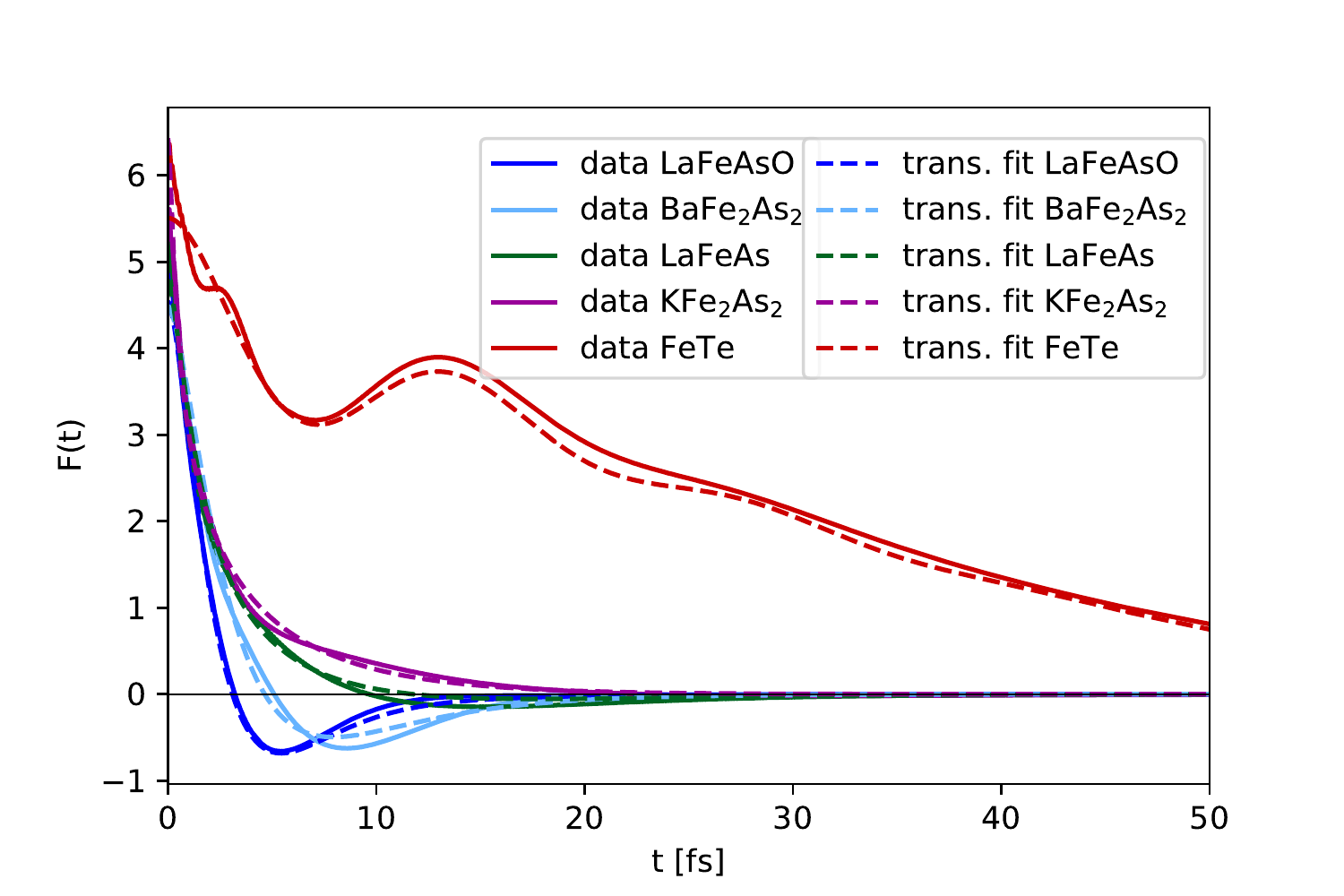}
	\endminipage
	\caption{ Dissipative part of the spin-spin susceptibility obtained by MaxEnt (left figure solid lines). The harmonic-oscillator fits are shown as dashed lines. Center figure: Spin-spin susceptibility in time. Direct transform of MaxEnt data shown as solid lines and the  analytic expression for the fitted model as dashed lines. Right figure: Spin-spin anti-commutator correlation function for data (solid lines) and fitted model (dashed lines).}\label{fig:fit_checks}
\end{figure}

\begin{table}[ht!]
	\centering
	\ra{1.3}
	\hspace*{-0.0cm}
	\vspace*{4.5mm}
	\begin{tabular}{@{}lllll@{}}\toprule
		& $\omega_0 $[eV]   & $\gamma$ [eV] &  $t_\gamma$ [fs] 	& $t_{\bar{\omega}}$ [fs]  \\
		\midrule
		LaFeAsO			& \textbf{0.39}	& 0.35 & 	1.9 		 	& 3.8     	\\
		BaFe$_2$As$_2$	& \textbf{0.28}    & \textbf{0.28} & 	2.4 		& 15.2		\\
		LiFeAs 			& 0.30    & \textbf{0.58} & 	7.9 		& -   		\\
		KFe$_2$As$_2$ 	& 0.51    & \textbf{2.08} &  10.3			& -	  	  	\\
		FeTe 			& \textbf{0.029}    & 0.022 &  29.3			& 34.8	  	\\
		\bottomrule
	\end{tabular}
	\caption{Fitting parameters extracted with a harmonic oscillator model (second and third column), effective lifetime $\chi(t\rightarrow \infty)\propto \mathrm{e}^{-t/t_\gamma} $ (third column) and effective oscillation frequency $t_{\bar{\omega}} = \frac{\hbar}{\sqrt{\omega_0^2 - \gamma^2}}$(fourth column)}
	\label{tab:fit_params}
\end{table}
%
%
%
%
One can also define a harmonic-oscillator anti-commutator through the fluctuation-dissipation theorem as 
$F(\omega) = \frac{1}{\pi} \coth(\omega \beta/2) \mathrm{Im}\chi^R(\omega) $. For the latter it is not easy to get an analytical expression for the Fourier-transform ($F^{\mathrm{harm. \, osz.}}(t)  = \int_{-\infty}^{\infty} \mathrm{d}\omega \, \mathrm{e}^{- \mathrm{i} \omega t}  \frac{1}{\pi} \coth(\omega \beta/2) $). Therefore we preformed the transformation only numerically. To assess the quality of the fit we checked also $\chi^{\mathrm{data}}(t)$ against the analytical expressions given in \cref{eqn:Analytical_chi_t}. The results of the transformation of the data as well as the transformation of the fits is shown in \cref{fig:fit_checks}. For LaFeAsO, BeFe$_2$As$_2$, LiFeAs and KFe$_2$As$_2$ a single peak model was used, while for FeTe the double-peak-structure in the data necessitated a two-peak model. Due to the second peak in FeTe no sign-change is observed in the corresponding $\chi(t)$, although the main (first) peak would predict an oscillatory (under-damped) behavior. 

A comparison between the right and the center-part of \cref{fig:fit_checks} shows the same behavior qualitatively, although quantitative differences are observed. One reason for the difference is the additional energy scale (temperature). 

{\it Comments on previous works -} 
The estimated values of the local fluctuating moment $\braket{m^2_{\text{loc}}}$ in the specific case of KFe$_2$As$_2$ obtained by our DFT+DMFT study deviates from the results of \cite{Wang2013}, where  $\braket{m^2}=0.1 \pm 0.02 $ [$\mu_B^2/\text{Fe}$]. According to our work KFe$_2$As$_2$ should not have a significantly different local magnetic moment compared with the other iron-pnictides/-chalcogenides, which is consistent with the more recent theoretical/experimental analysis of \cite{Lafuerza2017, Pelliciari2017}.

\bibliography{./TIMES}